\documentclass[10pt]{article}
\usepackage{latexsym}
\usepackage{amssymb}
\usepackage{amsmath}
\usepackage{amscd}
\usepackage{amsthm}
\usepackage[left=1.5cm,top=2cm,right=2cm,bottom=1.5cm]{geometry}

\usepackage[dvips]{graphicx}
\usepackage{hyperref}
\begin{document}
\begin{center}
\large{\bf {Cylindrically Symmetric Inhomogeneous Universes with a Cloud of Strings}} \\
\vspace{10mm}
\normalsize{Anil Kumar Yadav $^{1,\;\dag}$, Vineet Kumar Yadav $^{2a}$ and Lallan 
Yadav $^{2b}$}\\
\vspace{2mm}
\footnotetext[1]{Department of Physics, Anand Engineering
College, Keetham, Agra - 282 007, India.\;\; $^\dag$ E-mail: abanilyadav@yahoo.co.in} 
\vspace{5mm}
\footnotetext[2] {Department of Physics, D. D. U. Gorakhpur University, Gorakhpur - 273 009, India.} 
\end{center}
\vspace{4mm}
%\date{}
%\maketitle
\begin{abstract} 
Cylindrically symmetric inhomogeneous string cosmological models are investigated in presence of 
string fluid as a source of matter. To get the three types of exact solutions of Einstein's field 
equations we assume $A = f(x)k(t)$, $B = g(x)\ell(t)$ and $C = h(x)\ell(t)$. Some physical and 
geometric aspects of the models are discussed.   
\end{abstract}
%\newpage
%%%%%%%%%%%%%%%%%%%%%%%%%%%%%%%%%%%%%%%%%%%%%%%%%%%%%%%%%%%%%%%%%%%%%%%%%%%%%%%%%%%
%%%%%%%%%%%%%%%%%%%%%%%%%%%%%%%   SECTION 1  %%%%%%%%%%%%%%%%%%%%%%%%%%%%%%%%%%%%%%
\section{Introduction}
In recent years, there has been considerable interest in string cosmology because 
cosmic strings play an important role in the study of the early universe. These strings 
arise during the phase transition after the big bang explosion as the temperature goes 
down below some critical temperature as predicted by grand unified theories (Zel'dovich 
et al., 1975; Kibble, 1976, 1980; Everett, 1981; Vilenkin, 1981). Moreover, the 
investigation of cosmic strings and their physical processes near such strings has 
received wide attention because it is believed that cosmic strings give rise to 
density perturbations which lead to formation of galaxies (Zel'dovich, 1980; 
Vilenkin, 1981). These cosmic strings have stress energy and couple to the 
gravitational field. Therefore, it is interesting to study the gravitational effect 
which arises from strings by using Einstein's equations. \\

The general treatment of strings was initiated by Letelier (1979, 1983) and 
Stachel (1980). Letelier (1979) obtained the general solution of Einstein's 
field equations for a cloud of strings with spherical, plane and a particular 
case of cylindrical symmetry. Letelier (1983) also obtained massive string 
cosmological models in Bianchi type-I and Kantowski-Sachs space-times. Benerjee 
et al. (1990) have investigated an axially symmetric Bianchi type I string dust 
cosmological model in presence and absence of magnetic field. Exact solutions of 
string cosmology for Bianchi type-II, $-VI_{0}$, -VIII and -IX space-times have 
been studied by Krori et al. (1990) and Wang (2003). Wang (2004, 2005, 2006) and 
Yadav et al. (2007a,b) have investigated bulk viscous string cosmological models in 
different space-times. Bali et al. (2001, 2003, 2005, 2006, 2007) have obtained 
Bianchi type-I, -III, -V and type-IX string cosmological models in general 
relativity. The string cosmological models with a magnetic field are discussed by 
Chakraborty (1991), Tikekar and Patel (1992, 1994), Patel and Maharaj (1996). Ram 
and Singh (1995) obtained some new exact solution of string cosmology with and without a 
source free magnetic field for Bianchi type I space-time in the different basic form 
considered by Carminati and McIntosh (1980). Singh and Singh (1999) investigated string 
cosmological models with magnetic field in the context of space-time with $G_{3}$ 
symmetry. Singh (1995a,b) has studied string cosmology with electromagnetic fields in 
Bianchi type-II, -VIII and -IX space-times. Lidsey, Wands and Copeland (2000) have 
reviewed aspects of super string cosmology with the emphasis on the cosmological 
implications of duality symmetries in the theory. Yavuz et al. (2005) have examined 
charged strange quark matter attached to the string cloud in the spherical symmetric 
space-time admitting one-parameter group of conformal motion. Kaluza-Klein 
cosmological solutions are obtained by Yilmaz (2006) for quark matter attached to the 
string cloud in the context of general relativity. Recently Pradhan et al. (2007) have 
studied higher dimensional strange quark matter coupled to the string cloud with 
electromagnetic field admitting one parameter group of conformal motion.  \\   

Cylindrically symmetric space-time play an important role in the study of the universe 
on a scale in which anisotropy and inhomogeneity are not ignored. Inhomogeneous 
cylindrically symmetric cosmological models have significant contribution in 
understanding some essential features of the universe such as the formation of 
galaxies during the early stages of their evolution. Bali and Tyagi (1989) and 
Pradhan et al. (2001, 2006) have investigated cylindrically symmetric inhomogeneous 
cosmological models in presence of electromagnetic field. Barrow and Kunze (1997, 1998) 
found a wide class of exact cylindrically symmetric flat and open inhomogeneous string 
universes. In their solutions all physical quantities depend on at most one space 
coordinate and the time. The case of cylindrical symmetry is natural because of the 
mathematical simplicity of the field equations whenever there exists a direction in 
which the pressure equal to energy density. \\

Recently Baysal et al. (2001), Kilinc and Yavuz (1996), Pradhan (2007) and Pradhan et al. 
(2007a,b) have investigated some string cosmological models in cylindrically symmetric 
inhomogeneous universe. In this paper, we have revisited their solutions and obtained a 
new class of solutions. Here, we extend our understanding of inhomogeneous string cosmologies 
by investigating the simple models of non-linear cylindrically symmetric inhomogeneities 
outlined above. This paper is organized as follows: The metric and field equations are 
presented in Section $2$. In Section $3$, we deal with the solution of the field equations 
in three different cases. Finally, the results are discussed in Section $4$. The solutions 
obtained in this paper are new and different from the other author's solutions.  
%%%%%%%%%%%%%%%%%%%%%%%%%%%%%%%%%%%%%%%%%%%%%%%%%%%%%%%%%%%%%%%%%%%%%%%%%%%%%%%%%%%
%%%%%%%%%%%%%%%%%%%%%%%%%%%%%%%  SECTION 2  %%%%%%%%%%%%%%%%%%%%%%%%%%%%%%%%%%%%%%%%
\section{The Metric and Field  Equations}
We consider the metric in the form 
\begin{equation}
\label{eq1}
ds^{2} = A^{2}(dx^{2} - dt^{2}) + B^{2} dy^{2} + C^{2} dz^{2},
\end{equation}
where $A$, $B$ and $C$ are functions of $x$ and $t$.
The Einstein's field equations for a cloud of strings read as (Letelier, 1983) 
\begin{equation}
\label{eq2}
G^{j}_{i} \equiv  R^{j}_{i} - \frac{1}{2} R g^{j}_{i} = -(\rho u_{i}u^{j} - 
\lambda x_{i}x^{j}),
\end{equation}
where $u_{i}$ and $x_{i}$ satisfy conditions
\begin{equation}
\label{eq3}
u^{i} u_{i} = - x^{i} x_{i} = -1,
\end{equation}
and
\begin{equation}
\label{eq4}
u^{i} x_{i} = 0.
\end{equation}
Here, $\rho$ is the rest energy of the cloud of strings with massive particles 
attached to them. $\rho = \rho_{p} + \lambda$, $\rho_{p}$ being the rest energy 
density of particles attached to the strings and  $\lambda$ the density of tension 
that characterizes the strings. The unit space-like vector  $x^{i}$ represents 
the string direction in the cloud, i.e. the direction of anisotropy and the 
unit time-like vector  $u^{i}$ describes the four-velocity vector of the matter 
satisfying the following conditions 
\begin{equation}
\label{eq5}
g_{ij} u^{i} u^{j} = -1.
\end{equation}
In the present scenario, the comoving coordinates are taken as 
\begin{equation}
\label{eq6}
u^{i} = \left(0, 0, 0, \frac{1}{A}\right) 
\end{equation}
and choose $x^{i}$ parallel to x-axis so that
\begin{equation}
\label{eq7}
x^{i} = \left(\frac{1}{A}, 0, 0, 0 \right). 
\end{equation}
The Einstein's field equations (\ref{eq2}) for the line-element (\ref{eq1}) 
lead to the following system of equations:  
\[
G^{1}_{1} \equiv \frac{B_{44}}{B} + \frac{C_{44}}{C} - \frac{A_{4}}{A}
\left(\frac{B_{4}}{B} + \frac{C_{4}}{C}\right) - \frac{A_{1}}{A}
\left(\frac{B_{1}}{B} + \frac{C_{1}}{C}\right) -\frac{B_{1}C_{1}}{BC} + 
\frac{B_{4} C_{4}}{B C} 
\]
\begin{equation}
\label{eq8}
= \lambda A^{2},
\end{equation}
\begin{equation}
\label{eq9}
G^{2}_{2} \equiv \left(\frac{A_{4}}{A}\right)_{4} - \left(\frac{A_{1}}
{A}\right)_{1} + \frac{C_{44}}{C} - \frac{C_{11}}{ C} = 0,
\end{equation}
\begin{equation}
\label{eq10}
G^{3}_{3} \equiv \left(\frac{A_{4}}{A}\right)_{4} - \left(\frac{A_{1}}{A}\right)_{1} 
+ \frac{B_{44}}{B} - \frac{B_{11}}{B} =  0,
\end{equation}
\[
G^{4}_{4} \equiv - \frac{B_{11}}{B} - \frac{C_{11}}{C} + \frac{A_{1}}{A}
\left(\frac{B_{1}}{B} + \frac{C_{1}}{C}\right) + \frac{A_{4}}{A}\left(\frac{B_{4}}{B} 
+ \frac{C_{4}}{C}\right) - \frac{B_{1}C_{1}}{BC}  + \frac{B_{4} C_{4}}{B C} 
\]
\begin{equation}
\label{eq11}
= \rho A^{2},
\end{equation}
\begin{equation}
\label{eq12}
G^{1}_{4} \equiv \frac{B_{14}}{B} + \frac{C_{14}}{C} - \frac{A_{4}}{A}\left(\frac{B_{1}}{B}
 + \frac{C_{1}}{C}\right) - \frac{A_{1}}{A}\left(\frac{B_{4}}{B} + \frac{C_{4}}{C}\right) = 0,
\end{equation}
where the sub indices $1$ and $4$ in A, B, C and elsewhere denote differentiation
with respect to $x$ and $t$, respectively.

The velocity field $u^{i}$ is irrotational. The scalar expansion $\theta$, shear scalar 
$\sigma^{2}$, acceleration vector $\dot{u}_{i}$ and proper volume $V^{3}$ are respectively 
found to have the following expressions:
\begin{equation}
\label{eq13}
\theta = u^{i}_{;i} = \frac{1}{A}\left(\frac{A_{4}}{A} + \frac{B_{4}}{B} + \frac{C_{4}}{C}
\right),
\end{equation}
\begin{equation}
\label{eq14}
\sigma^{2} = \frac{1}{2} \sigma_{ij} \sigma^{ij} = \frac{1}{3}\theta^{2} - \frac{1}{A^{2}}
\left(\frac{A_{4}B_{4}}{AB} + \frac{B_{4}C_{4}}{BC} + \frac{C_{4}A_{4}}{CA}\right),
\end{equation}
\begin{equation}
\label{eq15}
\dot{u}_{i} = u_{i;j}u^{j} = \left(\frac{A_{1}}{A}, 0, 0, 0\right), 
\end{equation}
\begin{equation}
\label{eq16}
V^{3} = \sqrt{-g} = A^{2} B C,
\end{equation}
where $g$ is the determinant of the metric (\ref{eq1}). Using the field equations and 
the relations (\ref{eq13}) and (\ref{eq14}) one obtains the Raychaudhuri's equation as
\begin{equation}
\label{eq17}
\dot{\theta} = \dot{u}^{i}_{;i} - \frac{1}{3}\theta^{2} - 2 \sigma^{2} - \frac{1}{2} 
\rho_{p},
\end{equation}
where dot denotes differentiation with respect to $t$ and
\begin{equation}
\label{eq18}
R_{ij}u^{i}u^{j} = \frac{1}{2}\rho_{p}.
\end{equation}
 
With the help of equations (\ref{eq1}) - (\ref{eq7}), the Bianchi identity 
$\left(T^{ij}_{;j}\right)$ reduced to two equations:
\begin{equation}
\label{eq19}
\rho_{4} - \frac{A_{4}}{A}\lambda + \left(\frac{A_{4}}{A} + \frac{B_{4}}{B} + 
\frac{C_{4}}{C}\right)\rho = 0
\end{equation}
and
\begin{equation}
\label{eq20}
\lambda_{1} - \frac{A_{1}}{A}\rho + \left(\frac{A_{1}}{A} + \frac{B_{1}}{B} + 
\frac{C_{1}}{C}\right)\lambda = 0.
\end{equation}
Thus due to all the three (strong, weak and dominant) energy conditions, one finds 
$\rho \geq 0$ and $\rho_{p} \geq 0$, together with the fact that the sign of $\lambda$ 
is unrestricted, it may take values positive, negative or zero as well.  
%%%%%%%%%%%%%%%%%%%%%%%%%%%%%%%%%%%%%%%%%%%%%%%%%%%%%%%%%%%%%%%%%%%%%%%%%%%%%%%%%%%
%%%%%%%%%%%%%%%%%%%%%%%%%%%%%%%  SECTION 3  %%%%%%%%%%%%%%%%%%%%%%%%%%%%%%%%%%%%%%%%
\section{Solutions of the Field  Equations}
From Eqs. (\ref{eq9}) and (\ref{eq10}), we obtain
\begin{equation}
\label{eq21}
\frac{B_{44}}{B} - \frac{B_{11}}{B} = \frac{C_{44}}{C} - \frac{C_{11}}{C},
\end{equation}
and
\begin{equation}
\label{eq22}
2\left(\frac{A_{4}}{A}\right)_{4} - 2\left(\frac{A_{1}}{A}\right)_{1} + \frac{B_{44}}{B} + 
\frac{C_{44}}{C} - \frac{B_{11}}{B} - \frac{C_{11}}{C} = 0. 
\end{equation}
To get determinate solution we assume
$$ A = f(x)k(t), $$
$$ B = g(x)\ell(t), $$
\begin{equation}
\label{eq23}
C = h(x)\ell(t). 
\end{equation}
and
\begin{equation}
\label{eq24}
\frac{f_{1}}{f} = \mbox{m (constant)}.
\end{equation}
Using (\ref{eq23}) in (\ref{eq21}), we get
\begin{equation}
\label{eq25}
\frac{g_{11}}{g} = \frac{h_{11}}{h}.
\end{equation}
Equations (\ref{eq22}), (\ref{eq23}) and (\ref{eq25}), we have
\begin{equation}
\label{eq26}
\left(\frac{k_{44}}{k} - \frac{k^{2}_{4}}{k^{2}} + \frac{\ell_{44}}{\ell}\right) = 
\left(\frac{f_{11}}{f} - \frac{f^{2}_{1}}{f^{2}} + \frac{g_{11}}{g}\right) = \mbox{s (constant)}. 
\end{equation}
Using (\ref{eq24}) in right hand side of Eq. (\ref{eq26}) leads to
\begin{equation}
\label{eq27}
\frac{g_{11}}{g} = \mbox{s},
\end{equation}
which with the use of (\ref{eq25}) leads to
\begin{equation}
\label{eq28}
 \frac{g_{11}}{g} = \frac{h_{11}}{h} = \mbox{s}.
\end{equation}
Equation (\ref{eq28}) leads to
\begin{equation}
\label{eq29}
g = h =  \left[ \begin{array}{ll}
            c_{1} \cosh(\sqrt{s} x + x_{0})  & \mbox { when $s>0$}\\
 (c_{1} x + c_{2})                                        & \mbox { when $s=0$} \\
            c_{1}\cos(\sqrt{\mid s \mid} x + x_{0})           & \mbox { when $s<0$}
            \end{array} \right. 
\end{equation}
where $c_{1}$, $c_{2}$ and $x_{0}$ are constants of integration. \\
Using Eq. (\ref{eq23}) in (\ref{eq12}), we have
\begin{equation}
\label{eq30}
\frac{\frac{\ell_{4}}{\ell}}{\frac{k_{4}}{k}} = \frac{\frac{g_{1}}{g} + \frac{h_{1}}{h}}
{\frac{g_{1}}{g} + \frac{h_{1}}{h} - \frac{2f_{1}}{f}} = \mbox{b (constant)}.
\end{equation}
Equation (\ref{eq30}) leads to
\begin{equation}
\label{eq31}
\frac{\ell_{4}}{\ell} = b\frac{k_{4}}{k},
\end{equation}
which after integration gives
\begin{equation}
\label{eq32}
\ell = nk^{b},
\end{equation}
where $n$ is constant of integration. Using (\ref{eq32}) in \ref{eq26}), we have
\begin{equation}
\label{eq33}
\frac{\ell_{44}}{\ell} - \frac{1}{(1 + b)}\frac{\ell^{2}_{4}}{\ell^{2}} = N,
\end{equation}
where $N = \frac{sb}{(1 + b)}$. Equation (\ref{eq33}) leads to
\begin{equation}
\label{eq34}
\ell = \left[ \begin{array}{ll}
 c^{\frac{(1 + b)}{b}}_{3} \cosh^{\frac{(1 + b)}{b}}(\sqrt{r} t + t_{0})  & \mbox { when $r>0$}\\
 (c_{3} t + c_{4})^{\frac{(1 + b)}{b}}                                        & \mbox { when $r=0$} \\
            c^{\frac{(1 + b)}{b}}_{3}\cos^{\frac{(1 + b)}{b}}(\sqrt{\mid r \mid} t + t_{0})    & \mbox 
{ when $ r < 0$}
            \end{array} \right. 
\end{equation}
where $r = \frac{Nb}{(1 + b)}$ and $c_{3}$, $c_{4}$, $t_{0}$ are constants of integration. \\
Hence from (\ref{eq32}) we obtain
\begin{equation}
\label{eq35}
k = \left[ \begin{array}{ll}
 \frac{c^{\frac{(1 + b)}{b^{2}}}_{3}}{n^{\frac{1}{b}}} \cosh^{\frac{(1 + b)}{b^{2}}}(\sqrt{r} t + t_{0}) 
 & \mbox { when $r>0$}\\
 \frac{(c_{3} t + c_{4})^{\frac{(1 + b)}{b^{2}}}}{n^{\frac{1}{b}}}       & \mbox { when $r=0$} \\
            \frac{c_{3}^{\frac{(1 + b)}{b^{2}}}}{n^{\frac{1}{b}}}\cos^{\frac{(1 + b)}{b^{2}}}
(\sqrt{\mid r \mid } t + t_{0})           & \mbox { when $ r < 0$}
            \end{array} \right. 
\end{equation}
Eq. (\ref{eq24}) leads to
\begin{equation}
\label{eq36}
f = e^{mx}.
 \end{equation}
Now we consider the following three cases:
%%%%%%%%%%%%%%%%%%%%%%%%%%%%%%%%%%%%%%%%%%%%%%%%%%%%%%%%%%%%%%%%%%%%%%%%
%%%%%%%%%%%%%%%%%%%%%%%%%% SUBSECTION 3.1  %%%%%%%%%%%%%%%%%%%%%%%%%%%%%%%%%%
\subsection{When $s>0$ and $r > 0$}
In this case we have
\begin{equation}
\label{eq37}
A = \alpha e^{mx} \cosh^{\frac{(1 + b)}{b}}(\sqrt{r} t + t_{0}),
 \end{equation}
\begin{equation}
\label{eq38}
B = C = \beta \cosh(\sqrt{s}x + x_{0}) \cosh^{\frac{(1 + b)}{b}}(\sqrt{r} t + t_{0}),
\end{equation}
where $\alpha = \frac{c_{3}^{\frac{(1 + b)}{b^{2}}}}{n^{\frac{1}{b}}}$ and 
$\beta = c_{1} c_{3}^{\frac{(1 + b)}{b}}$. \\
After using suitable transformation of the co-ordinates 
\begin{equation}
\label{eq39}
X = x + \frac{x_{0}}{\sqrt{s}}, \, Y = y, \, Z = z, \, T = t + \frac{t_{0}}{\sqrt{r}},
\end{equation}
the metric (\ref{eq1}) reduces to the form
\[
ds^{2} =  \alpha^{2} e^{2m(X - \frac{x_{0}}{\sqrt{s}})} \cosh^{\frac{2(1 + b)}{b^{2}}}(\sqrt{r} T)
(dX^{2} - dT^{2}) + 
\]
\begin{equation}
\label{eq40}
+ \beta^{2} \cosh^{2}(\sqrt{s} X) \cosh^{\frac{2(1 + b)}{b}}(\sqrt{r} T) (dY^{2} + dZ^{2}).
\end{equation}
In this case the physical parameters, i.e. the string tension density $(\lambda)$, the energy 
density $(\rho)$, the particle density $(\rho_{p})$ and the kinematical parameters, i.e. the 
scalar of expansion $(\theta)$, shear tensor $(\sigma)$, the acceleration vector $(\dot{u}_{i})$ 
and the proper volume $(V^{3})$ for the model (\ref{eq40}) are given by  
\[
\lambda = \frac{1}{\alpha^{2} e^{2m(X - \frac{x_{0}}{\sqrt{s}})} \cosh^{\frac{2(1 + b)}{b^{2}}}
(\sqrt{r} T)}\Biggl[\frac{2(1 + b)r}{b^{2}} + \frac{(1 + b)(b^{2} + b -2)r}{b^{3}}\tanh^{2}(\sqrt{r} T)
\]
\begin{equation}
\label{eq41}
- \sqrt{s}\tanh(\sqrt{s}X)\{2me^{m(X - \frac{x_{0}}{\sqrt{s}})} + \sqrt{s}\tanh(\sqrt{s}X)\}\Biggr],
\end{equation}
\[
\rho = \frac{1}{\alpha^{2} e^{2m(X - \frac{x_{0}}{\sqrt{s}})} \cosh^{\frac{2(1 + b)}{b^{2}}}
(\sqrt{r} T)}\Biggl[\frac{(1 + b)^{2}(2 + b)r}{b^{3}}\tanh^{2}(\sqrt{r} T)
\]
\begin{equation}
\label{eq42}
+ \sqrt{s}\tanh(\sqrt{s}X)\{2me^{m(X - \frac{x_{0}}{\sqrt{s}})} - \sqrt{s}\tanh(\sqrt{s}X)\} - 2s\Biggr],
\end{equation}
\[
\rho_{p} = \frac{1}{\alpha^{2} e^{2m(X - \frac{x_{0}}{\sqrt{s}})} \cosh^{\frac{2(1 + b)}{b^{2}}}
(\sqrt{r} T)}\Biggl[\frac{2(1 + b)(2 + b)r}{b^{3}}\tanh^{2}(\sqrt{r} T)
\]
\begin{equation}
\label{eq43}
+ 4 \sqrt{s}e^{m(X - \frac{x_{0}}{\sqrt{s}})}\tanh(\sqrt{s}X) - \frac{2(1 + b)r}{b^{2}} - 2s\Biggr],
\end{equation}
\begin{equation}
\label{eq44}
\theta = \frac{(1 + b)(1 + 2b)\sqrt{r}\tanh(\sqrt{r} T)}{b^{2} \alpha e^{m(X - \frac{x_{0}}
{\sqrt{s}})} \cosh^{\frac{(1 + b)}{b^{2}}}(\sqrt{r} T)},
\end{equation}
\begin{equation}
\label{eq45}
\sigma^{2} = \frac{(1 + b)^{2}(b - 1)^{2} r \tanh^{2}(\sqrt{r} T)}{3b^{4} \alpha^{2} 
e^{2m(X - \frac{x_{0}}{\sqrt{s}})} \cosh^{\frac{2(1 + b)}{b^{2}}}(\sqrt{r} T)},
\end{equation}
\begin{equation}
\label{eq46}
\dot{u}_{i} = (m, 0, 0, 0),
\end{equation}
\begin{equation}
\label{eq47}
V^{3} = \sqrt{-g} = \alpha^{2} \beta^{2} e^{2m(X - \frac{x_{0}}{\sqrt{s}})} \cosh^{2}(\sqrt{s} X) 
\cosh^{\frac{2(1 + b)}{b^{2}}}(\sqrt{r} T).
\end{equation}
From (\ref{eq44}) and (\ref{eq45}), we obtain
\begin{equation}
\label{eq48}
\frac{\sigma^{2}}{\theta^{2}}=  \frac{(b - 1)^{2}}{3(2b + 1)^{2}}  = \mbox {constant}.
\end{equation}
The energy conditions $\rho \geq 0$ and $\rho_{p} \geq 0$ are satisfied for the model (\ref{eq40}). 
The conditions $\rho \geq 0$ and $\rho_{p} \geq 0$ lead to
\[
\frac{(1 + b)^{2}(2 + b)r}{b^{3}}\tanh^{2}(\sqrt{r}T) + \sqrt{s}\tanh(\sqrt{s}X)\times
\]
\begin{equation}
\label{eq49}
\left\{2me^{m(X - \frac{x_{0}}{\sqrt{s}})} - \sqrt{s}\tanh(\sqrt{s}X)\right\} \geq 2s.
\end{equation}
and
\[
\frac{2(1 + b)(2 + b)r}{b^{3}}\tanh^{2}(\sqrt{r}T) + 4\sqrt{s}e^{m(X - \frac{x_{0}}{\sqrt{s}})}
\tanh(\sqrt{s}X)
\]
\begin{equation}
\label{eq50}
\geq \frac{2(1 + b)r}{b^{2}} + 2s.
\end{equation}
respectively. From Eq. (\ref{eq41}), we observe that the string tension density $\lambda \geq 0$ 
leads to
\[
\tanh^{2}(\sqrt{r} T) \geq \frac{b^{3}\sqrt{s}\tanh(\sqrt{s}X)}{(1 + b)(b^{2} + b -2)}\left\{
2m e^{m(X - \frac{x_{0}}{\sqrt{s}})} + \sqrt{s}\tanh(\sqrt{s}X) \right\} 
\]
\begin{equation}
\label{eq51}
- \frac{2br}{(b^{2} + b - 2)}.
\end{equation}
The model (\ref{eq40}) are shearing, accelerating and non-rotating. For $b < -1$, when $T \to 0$, 
$\theta \to 0$ and when $T \to \infty $, $\theta \to \infty$. Hence for $b < -1$, the model is 
expanding. Since $\frac{\sigma}{\theta} =$ constant, therefore the model does not approach isotropy.
But we observe that for $b = 1$, shear scalar is zero and hence the model isotropizes for this value 
of b. We also observe that when $T \to \infty$, the proper volume $V^{3} \to \infty$ and $\rho \to 0$. 
Hence volume increases when $T$ increases and the energy density is a decreasing function of $T$. 
%%%%%%%%%%%%%%%%%%%%%%%%%%%%%%%%%%%%%%%%%%%%%%%%%%%%%%%%%%%%%%%%%%%%%%%%
%%%%%%%%%%%%%%%%%%%%%%%%%% SUBSECTION 3.2  %%%%%%%%%%%%%%%%%%%%%%%%%%%%%%%%%%
\subsection{When $s = 0$ and $r = 0$}
In this case we obtain
\begin{equation}
\label{eq52}
A = \frac{(c_{3}t + c_{4})^{\frac{(1 + b)}{b^{2}}}}{n^{\frac{1}{b}}} e^{mx} ,
\end{equation}
\begin{equation}
\label{eq53}
B = C = (c_{1}x + c_{2})(c_{3}t + c_{4})^{\frac{(1 + b)}{b}}.
\end{equation}
By using suitable transformation, the metric (\ref{eq1}) reduces to the form
\begin{equation}
\label{eq54}
ds^{2} = \alpha^{2} e^{2m(X - \delta)} T^{\frac{2(1 + b)}{b^{2}}} (dX^{2} - dT^{2}) + \beta^{2}X^{2}
T^{\frac{2(1 + b)}{b}}(dY^{2} + dZ^{2}),
\end{equation}
where $x = X - \delta$, $y = Y$, $z = Z$, $t + \frac{c_{4}}{c_{3}} = T$, $\delta = \frac{c_{2}}{c_{1}}$.
$\alpha$ and $\beta$ are already defined in previous section.\\
In this case the physical parameters, i.e. the string tension density $(\lambda)$, the energy 
density $(\rho)$, the particle density $(\rho_{p})$ and the kinematical parameters, i.e. the 
scalar of expansion $(\theta)$, shear tensor $(\sigma)$, the acceleration vector $(\dot{u}_{i})$ 
and the proper volume $(V^{3})$ for the model (\ref{eq40}) are given by 
\begin{equation}
\label{eq55}
\lambda = \frac{1}{ \alpha^{2} e^{2m(X - \delta)} T^{\frac{2(1 + b)}{b^{2}}}}\left[\frac{(1 + b)\{
b(3 + b) - 2(1 + b)\}}{b^{3}}\frac{1}{T^{2}} - \frac{2m}{X} - \frac{1}{X^{2}}\right],
\end{equation}
\begin{equation}
\label{eq56}
\rho = \frac{1}{ \alpha^{2} e^{2m(X - \delta)} T^{\frac{2(1 + b)}{b^{2}}}}\left[\frac{(1 + b)^{2}
(2 + b)}{b^{3}}\frac{1}{T^{2}} + \frac{2m}{X} - \frac{1}{X^{2}} \right],
\end{equation}
\begin{equation}
\label{eq57}
\rho_{p} = \frac{1}{ \alpha^{2} e^{2m(X - \delta)} T^{\frac{2(1 + b)}{b^{2}}}}\left[\frac{(1 + b)
(4 - b)}{b^{3}}\frac{1}{T^{2}} + \frac{4m}{X}\right],
\end{equation}
\begin{equation}
\label{eq58}
\theta = \frac{(1 + b)(1 + 2b)}{b^{2}}\frac{1}{ \alpha e^{m(X - \delta)} T^{\frac{b(1 + b) + 1}{b}}},
\end{equation}
\begin{equation}
\label{eq59}
\sigma^{2} = \frac{(1 + b)^{2}( b - 1)^{2}}{b^{3}}\frac{1}{ \alpha^{2} e^{2m(X - \delta)} T^{\frac{2b(1 + b) + 2}
{b^{2}}}},
\end{equation}
\begin{equation}
\label{eq60}
\dot{u}_{i} = (m, 0, 0, 0)
\end{equation}
\begin{equation}
\label{eq61}
V^{3} = \sqrt{-g} = \alpha^{2} \beta^{2} e^{2m(X - \delta)} X^{2} T^{\frac{2(1 + b)^{2}}{b^{2}}}
\end{equation}
\begin{equation}
\label{eq62}
\frac{\sigma^{2}}{\theta^{2}} = \frac{b(b - 1)^{2}}{(1 + 2b)^{2}} = \mbox{constant}.
\end{equation}
The energy conditions $\rho \geq 0$ and $\rho_{p} \geq 0$ are satisfied for the model (\ref{eq54}). 
The conditions $\rho \geq 0$ and $\rho_{p} \geq 0$ lead to
\begin{equation}
\label{eq63}
\frac{(1 + b)^{2}(2 + b)}{b^{3}}\frac{1}{T^{2}} \geq \frac{1}{X^{2}} - \frac{2m}{X^{2}},
\end{equation}
and
\begin{equation}
\label{eq64}
\frac{(1 + b)(4 - b)}{b^{3}}\frac{1}{T^{2}} + \frac{4m}{X} \geq 0,
\end{equation}
respectively. From Eq. (\ref{eq55}), we observe that the string tension density $\lambda \geq 0$ 
leads to
\begin{equation}
\label{eq65}
\frac{(1 + b)\{b(3 + b) - 2(1 + b)\}}{b^{3}} \frac{1}{T^{2}} \geq \frac{2m}{X} + \frac{1}{X^{2}}.
\end{equation}
The model (\ref{eq54}) has singularity at $T = 0$. The model starts with a big bang singularity 
at $T = 0$ and continue to expand till $T = \infty$. Since $\frac{\sigma}{\theta} =$ constant, therefore 
the model does not approach isotropy. But we observe that for $b = 1$, shear scalar is zero and hence 
the model isotropizes for this value of b. We also observe that when $T \to \infty$, the proper 
volume $V^{3} \to \infty$ and $\rho \to 0$. Hence volume increases when $T$ increases and the energy 
density is a decreasing function of $T$. Generally the model (\ref{eq54}) represents an expanding, 
shearing, accelerating and non-rotating universe. 
%%%%%%%%%%%%%%%%%%%%%%%%%%%%%%%%%%%%%%%%%%%%%%%%%%%%%%%%%%%%%%%%%%%%%%%%
%%%%%%%%%%%%%%%%%%%%%%%%%% SUBSECTION 3.3  %%%%%%%%%%%%%%%%%%%%%%%%%%%%%%%%%%
\subsection{When $s < 0$ and $r < 0$}
In this case we obtain
\begin{equation}
\label{eq66}
A = \alpha e^{m x}\cos^{\frac{(b + 1)}{b^{2}}}(\sqrt{\mid r \mid} t + t_{0}),
\end{equation}
\begin{equation}
\label{eq67}
B = C = \beta \cos(\sqrt{\mid s \mid} x + x_{0})\cos^{\frac{(b + 1)}{b}}(\sqrt{\mid r \mid)} t + t_{0}).
\end{equation}
By using suitable transformation, the metric (\ref{eq1}) reduces to the form
\[
ds^{2} = \alpha^{2} e^{2m(X - \frac{x_{0}}{\sqrt{s}})} \cos^{\frac{2(b + 1)}{b^{2}}}(\sqrt{\mid r \mid} T)
(dX^{2} - dT^{2}) \, + 
\]
\begin{equation}
\label{eq68}
\beta^{2}\cos^{2} (\sqrt{\mid s \mid} X) \cos^{\frac{2(b + 1)}{b^{2}}}(\sqrt{\mid r \mid} T)
(dY^{2} + dZ^{2}),
\end{equation}
In this case the physical parameters, i.e. the string tension density $(\lambda)$, the energy 
density $(\rho)$, the particle density $(\rho_{p})$ and the kinematical parameters, i.e. the 
scalar of expansion $(\theta)$, shear tensor $(\sigma)$, the acceleration vector $(\dot{u}_{i})$ 
and the proper volume $(V^{3})$ for the model (\ref{eq40}) are given by 
\[
\lambda = \frac{1}{\L^{2} e^{2m(X - \frac{x_{0}}{\sqrt{\mid s \mid}})} \cos^{\frac{2(b + 1)}{b^{2}}}
(\sqrt{\mid r \mid} T)}\times
\]
\[
\Biggl[- \frac{2(b + 1)\mid r \mid}{b} + \frac{(b + 1)(b^{2} + b - 2)}{b^{3}}\mid r \mid \tan^{2}(\sqrt{\mid r 
\mid} T)
\]
\begin{equation}
\label{eq69}
+ \, \sqrt{\mid s \mid} \tan(\sqrt{\mid s \mid} X)\left\{2n e^{m(X - \frac{x_{0}}{\sqrt{\mid s \mid}})} 
- \sqrt{\mid s \mid}\tan(\sqrt{\mid s \mid} X)\right\}\Biggr],
\end{equation}
where $\alpha = - \L$, $\L > 0$.
\[
\rho = \frac{1}{\L^{2} e^{2m(X - \frac{x_{0}}{\sqrt{\mid s \mid}})} \cos^{\frac{2(b + 1)}{b^{2}}}
(\sqrt{\mid r \mid} T)}\times
\]
\[
\Biggl[\frac{(b + 2)(b + 1)^{2}}{b^{3}}\mid r \mid \tan^{2}(\sqrt{\mid r \mid} T) + 2 \mid s \mid
\]
\begin{equation}
\label{eq70}
- \, \sqrt{\mid s \mid} \tan(\sqrt{\mid s \mid} X)\left\{2n e^{m(X - \frac{x_{0}}{\sqrt{\mid s \mid}})} 
+ \sqrt{\mid s \mid}\tan(\sqrt{\mid s \mid} X)\right\}\Biggr],
\end{equation}
\[
\rho_{p} = \frac{1}{\L^{2} e^{2m(X - \frac{x_{0}}{\sqrt{\mid s \mid}})} \cos^{\frac{2(b + 1)}{b^{2}}}
(\sqrt{\mid r \mid} T)}\times
\]
\[
\Biggl[\frac{2(b + 2)(b + 1)}{b^{3}}\mid r \mid \tan^{2}(\sqrt{\mid r \mid} T) + 2 \mid s \mid
\]
\begin{equation}
\label{eq71}
- \, 4\sqrt{\mid s \mid}m e^{m(X - \frac{x_{0}}{\sqrt{\mid s \mid}})}\tan(\sqrt{\mid s \mid} X)
+ \frac{2(b + 1) \mid r \mid }{b}\Biggr],
\end{equation}
\begin{equation}
\label{eq72}
\theta = \frac{(b + 1)(2b + 1)\mid r \mid \tan(\sqrt{\mid r \mid} T)}{b^{2} \L e^{m(X - \frac{x_{0}}
{\sqrt{\mid s \mid}})} \cos^{\frac{(b + 1)}{b^{2}}}(\sqrt{\mid r \mid} T),}         
\end{equation}
\begin{equation}
\label{eq73}
\sigma^{2} = \frac{(b - 1)^{2}(b + 1)^{2}\mid r \mid \tan^{2}(\sqrt{\mid r \mid} T)}{3b^{4}\L^{2} 
e^{2m(X - \frac{x_{0}}{\sqrt{\mid s \mid}})} \cos^{\frac{2(b + 1)}{b^{2}}}(\sqrt{\mid r \mid} T)},         
\end{equation}
\begin{equation}
\label{eq74}
\dot{u}_{i} = (m, 0, 0, 0),
\end{equation}
\begin{equation}
\label{eq75}
V^{3} = \L^{2} \beta^{2} e^{2m(X - \frac{x_{0}}{\sqrt{\mid s \mid}})} \cos^{2}(\sqrt{\mid s \mid} X). 
\cos^{\frac{2(b + 1)}{b^{2}}}(\sqrt{\mid r \mid} T) 
\end{equation}
From Eqs. (\ref{eq72}) and (\ref{eq73}), we obtain
\begin{equation}
\label{eq76}
\frac{\sigma^{2}}{\theta^{2}} = \frac{(b - 1)^{2}}{3(2b + 1)^{2}} = \mbox{constant}
\end{equation}
The model (\ref{eq68}) starts expanding at $T = 0$ and attains its maximum value at 
$T = \frac{\pi}{4\sqrt{\mid r \mid }}$. After that $\theta$ decreases to attain its minimum value at 
$T = \frac{3\pi}{4\sqrt{\mid r \mid }}$. The model oscillates with period $\frac{\pi}{2\sqrt{\mid r \mid}}$. 
The model is shearing and non-rotating. Since $\frac{\sigma}{\theta} =$ constant, therefore 
the model does not approach isotropy. But we observe that for $b = 1$, shear scalar vanishes. Hence 
$b = 1$ is the isotropy condition. 
%%%%%%%%%%%%%%%%%%%%%%%%%%%%%%%%%%%%%%%%%%%%%%%%%%%%%%%%%%%%%%%%%%%%%%%%%%%%%%%%%%%%%%%%%%%%%%%%%%%%%%%%%%%%%
%%%%%%%%%%%%%%%%%%%%%%%%%%  SECTION 4  %%%%%%%%%%%%%%%%%%%%%%%%%%%%%%%%%%%%%%%%%%%%%%%%%%%%%%%%%%%%%%%%%%%%%
\section{Concluding Remarks}
In this paper we have obtained a new class of exact solutions of Einstein's field equations in 
cylindrically symmetric inhomogeneous space-time with string fluid. The models (\ref{eq40}) and 
(\ref{eq54}), in general, are expanding, shearing and non-rotating. The model (\ref{eq68}) is 
oscillating, shearing and non-rotating. All these three models do not isotropize. 
The model (\ref{eq54}) is singular whereas the model (\ref{eq40}) is non-singular. It is important 
to note here that the models (\ref{eq40}), (\ref{eq54}) and (\ref{eq68}) reduce to homogeneous 
universe when $m = 0$ and $s = 0$. This shows that for $m = 0$ and $s = 0$, inhomogeneity dies out. 
In this case all these models are non-accelerating.\\
\section*{Acknowledgements} 
The authors thank the Harish-Chandra Research Institute, Allahabad, India 
for providing facility where this work was carried out.
%\newline
%\nonumsection{References}


\begin{thebibliography}{000}
\bibitem {ref1}
Bali, R. and Anjali: Astrophys. Space Sci. {\bf 302}, 201 (2006) 
\bibitem {ref2}
Bali, R. and Dave, S.: Astrophys. Space Sci. {\bf 288}, 503 (2001)
\bibitem {ref3}
Bali, R. and Singh, D.K.: Astrophys. Space Sci. {\bf 300}, 387 (2005)
\bibitem {ref4}
Bali, R. and Pradhan, A.: Chin. Phys. Lett. {\bf 24}, 585 (2007)
\bibitem {ref5}
Bali R., Pareek, U. K. and Pradhan, A.: Chin. Phys. Lett. {\bf 24}, 2455 (2007)
\bibitem {ref6}
Bali, R. and Tyagi, A.: Gen. Rel. Grav. {\bf 21}, 797 (1989)
\bibitem {ref7}
Bali, R. and Upadhaya, R.D.: Astrophys. Space Sci. {\bf 283}, 97 (2003)
\bibitem {ref8}
Banerjee, A., Sanyal, A.K. and Chakraborty, S.: Pramana-J. Phys. {\bf 34}, 1 (1990)
\bibitem {ref9}
Barrow, J.D. and Kunze, K.E.: Phys. Rev. D {\bf 56}, 741 (1997)
\bibitem {ref10}
Barrow, J.D. and Kunze, K.E.: Phys. Rev. D {\bf 57}, 2255 (1998)
\bibitem {ref11}
Baysal, H., Yavuz, I., Tarhan, I., Camci, U. and Yilmaz, I.: Turk. J. Phys. {\bf 25}, 
283 (2001)
\bibitem {ref12}
Carminati, J. and McIntosh, C.B.G.: J. Phys. A: Math. Gen. {\bf 13}, 953 (1980)
\bibitem {ref13}
Chakraborty, S.: Ind. J. Pure Appl. Phys. {\bf 29}, 31 (1980)
\bibitem {ref14}
Everett, A.E.: Phys. Rev. {\bf 24}, 858 (1981)
\bibitem {ref15}
Kibble, T.W.B.: J. Phys. A: Math. Gen. {\bf 9}, 1387 (1976)
\bibitem {ref16}  
Kibble, T.W.B.: Phys. Rep. {\bf 67}, 183 (1980)
\bibitem {ref17}
Kilinc, C.B. and Yavuz, I.: Astrophys. Space Sci. {\bf 238}, 239 (1996)
\bibitem {ref18}
Krori, K.D., Chaudhury, T., Mahanta, C.R. and Mazumder, A.: Gen. Rel. Grav. 
{\bf 22}, 123 (1990)
\bibitem {ref19}
Lidsey, J.E., Wands, D. and Copeland, E. J.: Phys. Rep. {\bf 337}, 343 (2000)
\bibitem {ref20}
Letelier, P.S.: Phys. Rev. D {\bf 20}, 1249 (1979) 
\bibitem {ref21}
Letelier, P.S.: Phys. Rev. D {\bf 28}, 2414 (1983)
\bibitem {ref22}
Patel, L.K. and Maharaj, S.D.: Pramana-J. Phys. {\bf 47}, 1 (1996)
\bibitem {ref23}  
Pradhan, A., Chakrabarty, I. and Saste, N. N.: Int. J. Mod. Phys. D 
{\bf 10}, 741 (2001)
\bibitem {ref24}  
Pradhan, A., Singh, P.K. and Jotania, K.: Czech. J. Phys. {\bf 56}, 641 (2006)
\bibitem {ref25}
Pradhan, A., Khadekar, G. S., Mishra, M. K. and Kumbhare, S.: Chin. Phys. Lett., {\bf 24}, 
3013 (2007).
\bibitem {ref26}
Pradhan, A., Yadav, A. K. and Singh, R. P. and Singh, V. K.: Astrophys. Space Sci, {\bf 312}, 
145 (2007a)
\bibitem {ref27}
Pradhan, A., Mishra, M.K. and Yadav, A.K.: gr-qc/0705.1765 (2007b)
\bibitem {ref28}
Pradhan, A.: Fizika B (Zagreb), {\bf 16}, 205 (2007).
\bibitem {ref29}
Ram, S. and Singh, T.K.: Gen. Rel. Grav. {\bf 27}, 1207 (1995)
\bibitem {ref30}
Singh, G.P. and Singh, T.: Gen. Rel. Grav. {\bf 31}, 371 (1999)
\bibitem {ref31}
Singh, G.P.: Nuovo Cim. B {\bf 110}, 1463 (1995a)
\bibitem {ref32}
Singh, G.P.: Pramana-J. Phys. {\bf 45}, 189 (1995b)
\bibitem {ref33}
Stachel, J.: Phys. Rev. D {\bf 21}, 2171 (1980)
\bibitem {ref34}
Tikekar, R. and Patel, L.K.: {\it Gen. Rel. Grav.} {\bf 24}, 397; (1994) (1992)
\bibitem {ref35} 
Tikekar, R. and Patel, L.K.: Pramana-J. Phys. {\bf 42}, 483 (1994) 
\bibitem {ref36}
Vilenkin, A.: Phys. Rev. D {\bf 24}, 2082 (1981)
\bibitem {ref37}
Wang, X.X.: Chin. Phys. Lett. {\bf 20}, 615 (2003)
\bibitem {ref38}
Wang, X.X.: Astrophys. Space Sci. {\bf 293}, 933 (2004)
\bibitem {ref39}
Wang, X.X.: Chin. Phys. Lett. {\bf 21}, 1205 (2004)
\bibitem {ref40}
Wang, X.X.: Chin. Phys. Lett. {\bf 22}, 29 (2005)
\bibitem {ref41}
Wang, X.X.: Chin. Phys. Lett. {\bf 23}, 1702 (2006)
\bibitem {ref42}
Yadav, M. K., Rai, A. and Pradhan, A.: Int. J. Theor. Phys. {\bf 46}, 2677 (2007a)
\bibitem {ref43}
Yadav, M. K., Pradhan, A. and Singh, S. K.: Astrophys. Space Sci., {\bf 311}, 423 (2007b)
\bibitem {ref44}
Yavuz, I., Yilmaz, I. and Baysal, H.: Int. J. Mod. Phys. D {\bf 14}, 1365 (2005)
\bibitem {ref45}
Yilmaz, I.: Gen. Rel. Grav. {\bf 38}, 1397 (2006)
\bibitem {ref46}
Zel'dovich, Ya. B.: Mon. Not. R. Astron. Soc. {\bf 192}, 663 (1980)
\bibitem {ref47}
Zel'dovich, Ya.B., Kobzarev, I.Yu. and Okun, L.B.: Zh. Eksp. Teor. Fiz. 
{\bf 67}, 3 (1975) 
\bibitem{ref48}
Zel'dovich, Ya.B., Kobzarev, I.Yu. and Okun, L.B.:Sov. Phys.-JETP {\bf 40}, 1 (1975)  
\end{thebibliography}
\end{document}